\documentclass[aip,apl,reprint,floatfix,citeautoscript,superscriptaddress]{revtex4-1}
\usepackage{graphicx}
\usepackage{bm,amsmath,amssymb,mathrsfs,dcolumn}
\usepackage[colorlinks=true, linkcolor=blue, citecolor=blue, urlcolor=blue, linktoc=page, bookmarks=false, pdfstartview={FitH}, pdfborder={0 0 0.0 [3 3]}]{hyperref} % HyperLink
\usepackage[usenames]{xcolor} 
\hypersetup{pdfborder=0 0 0,colorlinks=true,citecolor=blue,linkcolor=blue}

\begin{document}
\title{Recurrent Neural Networks Made of Magnetic Tunnel Junctions}
\author{Qi Zheng}
\author{Xiaorui Zhu}
\affiliation{Center for Advanced Quantum Studies and Department of Physics, Beijing Normal University, Beijing 100875, China}
\author{Yuanyuan Mi}
\affiliation{Center for Neurointelligence, Chongqing University, Chongqing 400044, China}
\author{Zhe Yuan}
\email{zyuan@bnu.edu.cn}
\affiliation{Center for Advanced Quantum Studies and Department of Physics, Beijing Normal University, Beijing 100875, China}
\affiliation{Center for Quantum Computing, Peng Cheng Laboratory, Shenzhen 518005, China}
\author{Ke Xia}
\affiliation{Center for Advanced Quantum Studies and Department of Physics, Beijing Normal University, Beijing 100875, China}
\affiliation{Center for Quantum Computing, Peng Cheng Laboratory, Shenzhen 518005, China}
\affiliation{Shenzhen Institute for Quantum Science and Engineering and Department of Physics, Southern University of Science and Technology, Shenzhen 518055, China}
\date{\today}
\begin{abstract}
Artificial intelligence based on artificial neural networks, which are originally inspired by the biological architectures of human brain, has mostly been realized using software but executed on conventional von Neumann computers, where the so-called von Neumann bottleneck essentially limits the executive efficiency due to the separate computing and storage units. Therefore, a suitable hardware platform that can exploit all the advantages of brain-inspired computing is highly desirable. Based upon micromagnetic simulation of the magnetization dynamics, we demonstrate theoretically and numerically that recurrent neural networks consisting of as few as 40 magnetic tunnel junctions can generate and recognize periodic time series after they are trained with an efficient algorithm. 
%With ultrahigh operating speed, nonvolatile memory and high endurance and reproducibility, spintronic devices are promising hardware candidates for neuromorphic computing. 
\end{abstract}
\maketitle

%%%%%%%%10%%%%%%%%20%%%%%%%%30%%%%%%%%40%%%%%%%%50%%%%%%%%60%%%%%%%%70%%%%%%%%80
%\section{Introduction}
In the past decade, significant progress has been made in artificial intelligence, where advanced algorithms using artificial neural networks (ANNs) have been successfully applied in image recognition, data classification, and other areas \cite{Lawrence97,Najafabadi15}. As an impressive example, the deep learning technique has shown an overwhelming advantage in the confrontation between a human and computer in the game of go \cite{Lecun15,Schmidhuber15,Silver16}. ANNs resulting from the simulating biological architectures of human brain possess the intrinsic advantages of brain including parallel computation, distributed storage, low energy consumption etc. Nevertheless, these advanced algorithms are mostly implemented using software and are still executed on conventional computers with the von Neumann architecture, where the advantages of brain-inspired computing are unfortunately not fully exploited \cite{Silver16}.

There have been many attempts to design and fabricate neuromorphic hardware devices \cite{Benjamin14,Neftci13,Davies18,Pei19}, which are not limited by the von Neumann bottleneck and intrinsically possess all the aforementioned advantages. Neuromorphic chips using the standard CMOS circuits such as the IBM TrueNorth consisting of billions of transistors can perform brain-inspired computing with remarkably low power \cite{Merolla14,Akopyan15}. Magnetic materials, however, have the potential to further increase the energy efficiency and areal density of devices by several orders of magnitude. An example has been shown in the devices of random number generation, where the most energy-efficient implementation of CMOS circuit consumes 2.9~pJ/bit and the circuit area of 4004 $\mu \mathrm m^2$~\cite{Mathew:ieeejssc12}. The device based on magnetic tunnel junctions (MTJs) only costs 20 fJ/bit and 2~$\mu\mathrm m^2$ in area~\cite{Vodenicarevic:prappl17}. In addition, memristors made of resistive and phase-change materials have attracted much attention in the realization of ANNs \cite{Wang18,Eryilmaz14,Suri11,Yang18,Yao16}. Compared to resistive and phase-change memristors, magnetic materials have faster dynamics at a time scale of nanoseconds and high endurance of more than $10^{15}$ cycles for magnetization switching \cite{Srinivasan16,Grollier16,Makarov11}. More importantly, the magnetization dynamics can be well described by the phenomenological Landau-Lifshitz-Gilbert equation \cite{Brown63,Gilbert04}, which has been examined in the past half century in the research communities of magnetism and spintronics. Recently, spintronics-based brain-inspired computing was used to realize the Hopfield model of memory \cite{Borders17}. The sound recognition could be significantly improved under the help of spin-torque nano-oscillators \cite{Romera18,Torrejon17}, whose nonlinear magnetization dynamics with memory is essential to capture the distinct acoustic features encoded in frequencies. A voltage-controlled stochastic spintronic device is implemented in experiment, where the stochastic behavior of magnetic switching is incorporated in an ANN to recognize the handwritten digits~\cite{cai:prappl19}. To date, most of the spintronic devices of brain-inspired computing have been applied in the recognition of static images or patterns, and little is known about their capability of temporal signal processing.

Reservoir computing is particularly suitable for encoding time series \cite{Jiang16,Li17}, in which the reservoir is physically a recurrent neural network (RNN) \cite{Lukosevicius:csr09}. The sparse and usually random connections among the neurons in the RNN ensure the capability to describe sufficiently complex functions \cite{Maass:neuralcomp02}. The relatively simple structure is another advantage of the RNN in the hardware implementation \cite{Furuta:prappl18,Jiang:arxiv19}. In this paper, we report a spintronic realization of RNNs with MTJs, which were used as the basic units of spin-transfer-torque magnetic random access memory. The nonlinear magnetization dynamics of an MTJ driven by an electrical current allows us to replace one neuron in the RNN by a single MTJ. By performing a micromagnetic simulation, we demonstrate that an RNN consisting of as few as 40 MTJs can generate and recognize sequential signals after an efficient training process. The capability of the network can be significantly improved by increasing the number of MTJs.

%%%%%%%%10%%%%%%%%20%%%%%%%%30%%%%%%%%40%%%%%%%%50%%%%%%%%60%%%%%%%%70%%%%%%%%80
%\section{MTJs as Artificial Neurons}

\begin{figure}[t]
\includegraphics[width=.85\columnwidth]{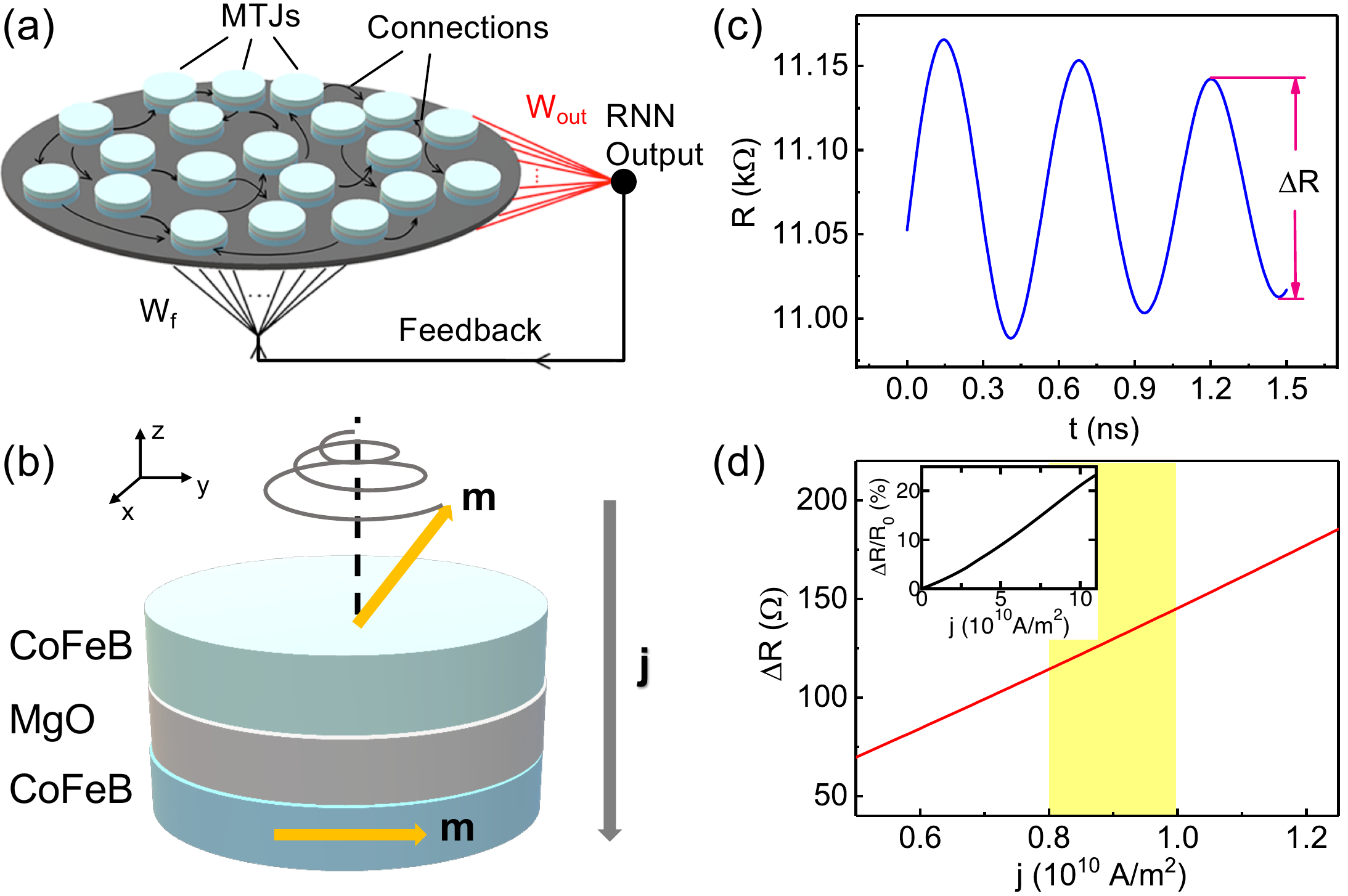}
\caption{(a) Sketch of an RNN made of MTJs. The black arrows denote the synaptic connections between MTJs. The red lines show the connections from every MTJ to the output node with adjustable weights. The black lines at the bottom represent the feedback connections that transport the output signal to every MTJ in the RNN. (b) The structure of an MTJ. (c) Damped oscillation in the resistance of the MTJ as a function of time due to the precessional motion of the free magnetic layer under the current density $j=0.9\times10^{10}$~A/m$^2$. The difference between the last maximum and minimum resistances $\Delta R$ within 1.5~ns after injecting the current is defined as the output. (d) The output of the artificial neuron as a function of current density. The shaded range of the current density is used to excite the magnetization dynamics. Inset: Calculated $\Delta R/R_0$ for a large range of $j$, where $R_0$ denotes the resistance of the MTJ at equilibrium.}\label{fig:1}
\end{figure}
We consider 40 MTJs as artificial neurons, which are randomly and sparsely connected with one another via either positive or negative unidirectional synapses, as schematically illustrated in Fig.~\ref{fig:1}(a). An MTJ consists of two thin magnetic layers separated by an insulating layer, as plotted in Fig.~\ref{fig:1}(b). The bottom layer has a fixed in-plane magnetization, usually pinned by an antiferromagnetic material via exchange bias \cite{Parkin99}. The magnetization of the top (free) layer in the MTJ can be excited by an injected electrical current following the Landau-Lifshitz-Gilbert equation in the presence of current-induced spin-transfer torques \cite{Slonczewski96,Berger96,Zhang02}
\begin{eqnarray}
\dot{\mathbf m}&=&-\gamma\mathbf m\times\mathbf H_{\rm eff}+\alpha\mathbf m\times\dot{\mathbf m}\nonumber\\
&&+\tau\epsilon\mathbf m\times\mathbf m_p\times\mathbf m-\beta\tau\mathbf m\times\mathbf m_p.\label{eq:llg}
\end{eqnarray}
Here, $\mathbf m$ is the magnetization direction of the free layer, and $\mathbf H_{\rm eff}$ is the effective magnetic field, including the exchange, anisotropic and demagnetization fields. The gyromagnetic ratio $\gamma$ and Gilbert damping $\alpha$ are both material parameters. The last two terms in Eq.~\eqref{eq:llg} are the adiabatic and nonadiabatic spin-transfer torques, respectively, where $\mathbf m_p$ denotes the magnetization direction of the fixed magnetic layer and the magnitude of the torque $\tau=(\gamma\hbar P/\mu_0 eM_s t)j$ depends on the current polarization $P$, the current density $j$, the saturation magnetization $M_s$ and the free-layer thickness $t$. The Slonczewski parameter $\epsilon=\Lambda^2/[(\Lambda^2+1)+(\Lambda^2-1)\mathbf m\cdot\mathbf m_p]$ characterizes the angular dependence of the torque with the dimensionless parameter $\Lambda\in[0,\,1]$. $\beta$ is the nonadiabaticity of the spin-transfer torque and is usually much smaller than one. In this work, the dynamic equation is solved numerically using the micromagnetic simulation program MuMax3 \cite{Vansteenkiste14}.

The magnetization of the free layer, which is perpendicular to the fixed layer at equilibrium, starts to precess about its easy axis with an external current. Therefore, the total resistance $R$ of the MTJ, which depends on the relative magnetization orientation of the two magnetic layers, exhibits an oscillation as a function of time. In the regime of a small current density, the amplitude of the oscillation decays gradually [see Fig.~\ref{fig:1}(c)], and the output signal of the artificial neuron is quantitatively defined by the difference between the last maximum and minimum values of the resistance ($\Delta R$) within 1.5~ns since the electrical current is injected.

In this way, the driving force of the magnetization precession, i.e., the injected electrical current density, can be defined as the input of the artificial neuron, while the resulting oscillatory resistance $\Delta R$ corresponds to the response or output. If one increases the current density, $\Delta R$ increases monotonically and nonlinearly. Using micromagnetic simulation, we can determine the nonlinear response function of the artificial neuron, which is plotted in Fig.~\ref{fig:1}(d). We choose an electrical current density in the range of $[0.8,\,1.0]\times10^{10} $~A/m$^2$ and a corresponding $\Delta R\in[114,\,145]~\Omega$ in this work. Both the input and output are normalized to be in the range $[0,\,1]$ when we consider the signal transfer among MTJs; see Supplementary Material. Owing to the small range of $j$ that we choose in this work, the resistance change $\Delta R$ is very small compared with the value at equilibrium $R_0=11.05$~k$\Omega$. In practice, $\Delta R/R_0$ can be increased up to 20\% by applying larger current density, as shown in the inset of Fig.~\ref{fig:1}(d).

Every MTJ is connected to the ``output neuron'' of the RNN via two synapses: one transfers the output signal of every MTJ to the ``output neuron'' ($w_{\rm out}$), and the other provides feedback from the ``output neuron'' to the MTJ ($w_{\rm f}$). Here, only the weights $w_{\rm out}$ are varied during the ``learning'' process, while the synapses within the RNN and the feedback synapses $w_{\rm f}$ are all fixed. Such an RNN can maintain time-dependent activation by the mutual interactions of neurons even without an external input. The detailed parameters and learning algorithm of this network can be found in Supplementary Material.

%\section{Generating Sinusoidal Functions}

\begin{figure}[t]
\includegraphics[width=.85\columnwidth]{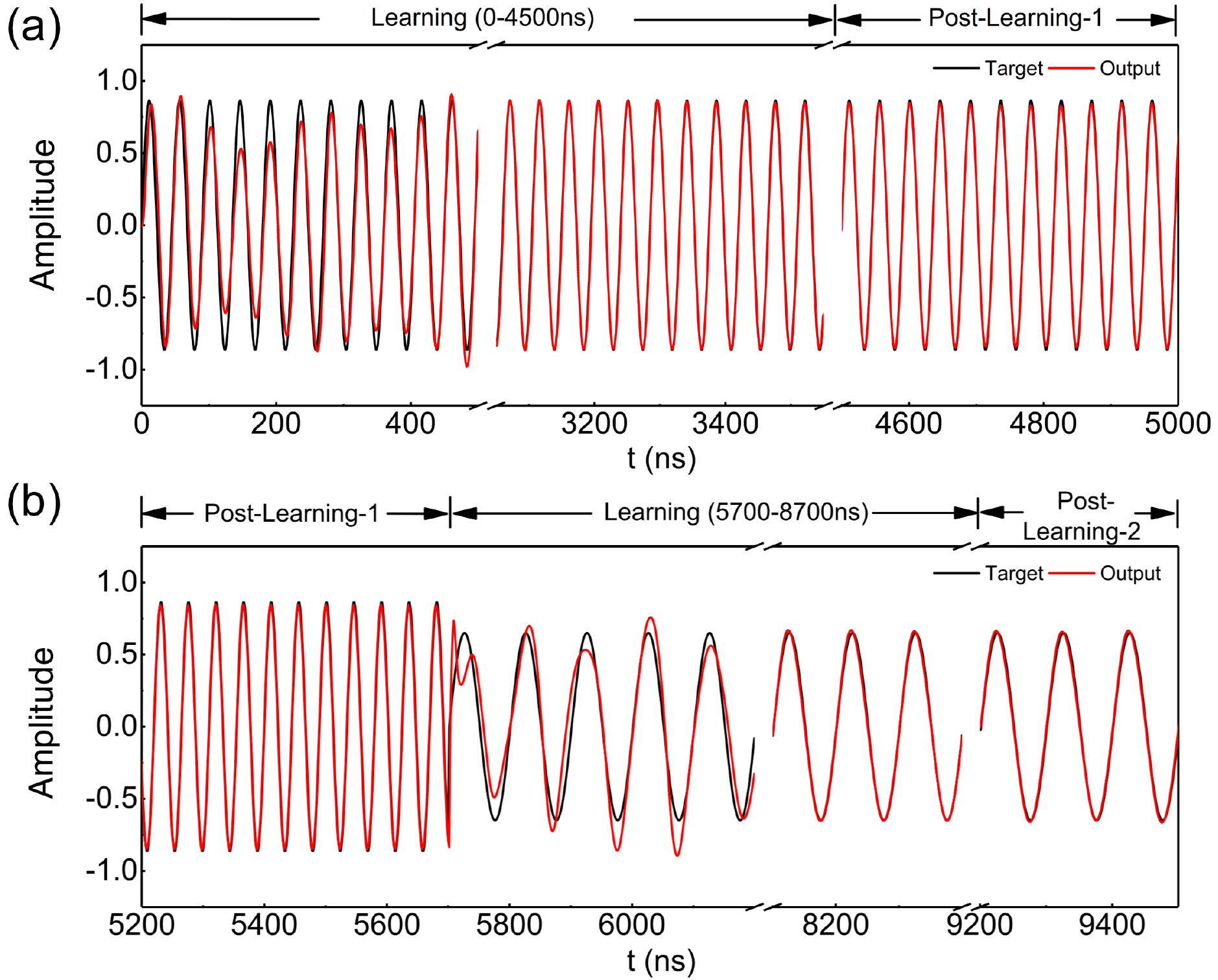}
\caption{(a) RNN output as a function of time (the red line). The target function is shown as the black line for comparison. Learning is performed in the first 4500~ns. (b) RNN output with a different target function after $t=5700$~ns.}\label{fig:2}
\end{figure}
As illustrated in Fig.~\ref{fig:1}(a), the weighted summation over the output signals of all the MTJs is defined as the output of the RNN. 
%Before ``learning'', the spontaneous output is a random function, since the initial connecting weights are essentially random. 
We first let the RNN generate a target sinusoidal function $f(t)=A\sin (2 \pi t/T)$ with $A=0.87$ and $T=45$~ns. A very efficient algorithm called ``force-learning scheme'' \cite{Sussillo09} is applied, in which the weights $w_{\rm out}$ are tuned by comparing the error between the RNN output and the desired target function. In practice, the RNN output follows the target function very quickly. As shown in Fig.~\ref{fig:2}(a), after 3000~ns, the output is already in perfect agreement with the target function. After learning for 4500~ns, we no longer vary $w_{\rm out}$, and the network sustains the generation of the same function as its output. This suggests the success of the learning scheme in this artificial RNN, which merely consists of 40 MTJs.

At $t=5700$~ns, we abruptly change the amplitude and period of the target sinusoidal function with $A=0.65$ and $T=80$~ns. At the same time, the force-learning algorithm is launched again to train the network via tuning the weights $w_{\rm out}$. The output significantly deviates from the new target function immediately after 5700~ns, but they superpose each other after 3000~ns of learning. We turn off the learning process after $t=8700$~ns, and the RNN steadily generates the new sinusoidal function afterwards.

%\section{Writing a Chinese Character}

\begin{figure}[t]
\includegraphics[width=.85\columnwidth]{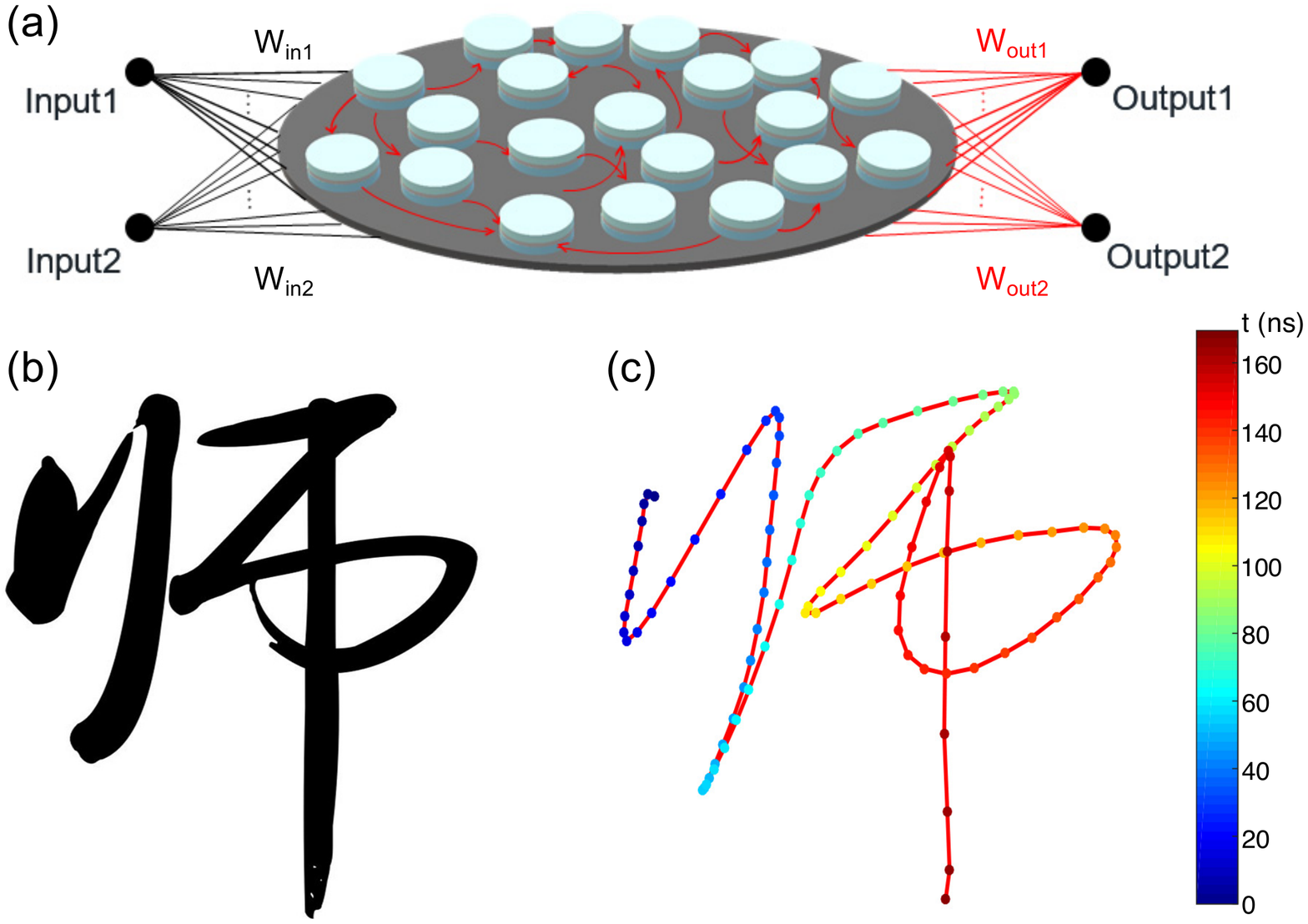}
\caption{(a) Sketch of the RNN used for writing a Chinese character with two input and two output nodes. The nodes output1 and output2 are used to generate the $x$ and $y$ coordinates, respectively, as a function of time. The connection weights among the MTJs and the output weights $w_{\rm out1}$ and $w_{\rm out2}$ are adjusted during the learning, which are illustrated by red lines. (b) The Chinese character meaning ``teacher'' written with a Chinese writing brush. (c) The output by the network in (a) reproducing the character in (b). Time is represented by colored dots with a uniform time interval of 1.5~ns.}
\label{fig:3}
\end{figure}

More complex time series can be learned using an RNN with more MTJs. For instance, by defining two-dimensional coordinates $x$ and $y$, which are both time-dependent functions, we can reproduce handwritten Chinese characters. As schematically illustrated in Fig.~\ref{fig:3}(a), we construct an RNN with 800 MTJs and two output nodes for $x$ and $y$. Instead of using feedback from the output nodes, we introduce two input nodes, where the ideal target functions are imported to the RNN to increase the learning efficiency. Moreover, we allow tunability of the random and sparse connections in the RNN to improve its flexibility and transferability, because a single RNN is used to generate the two coordinates simultaneously.

We choose the Chinese character meaning ``teacher'' written with a writing brush, as shown in Fig.~\ref{fig:3}(b). Two functions of time $x(t)$ and $y(t)$ are defined in a two-dimensional coordinate system to follow the stroke order of this character. Since both the connections in the RNN and the output weights are adjusted in the learning process, we employed the so-called innate training algorithm \cite{Laje13}. Such an algorithm is more robust and efficient for convergence. In addition, the innate training algorithm is practically highly resistant to noise or perturbations. The specific implementation of the innate training algorithm has two steps. In the first step, the connection weights inside the RNN are tuned to allow every MTJ to have its own sustained response to a pulse input. The success of this training is achieved when this sustained response becomes invariant for different initial conditions of the MTJs. This step is essential to improve the robustness of the network and produce insensitivity to noise.

Having adjusted the connection weights in the RNN, next we apply the force-learning algorithm to tune the output weights $w_{\rm out1}$ and $w_{\rm out2}$. In this step, the target periodic functions $x(t)$ and $y(t)$, which are implemented with the period of 170~ns, are imported from the two input nodes. After training for 10 periods, the output is plotted in Fig.~\ref{fig:3}(c), which successfully reproduces the handwritten Chinese character.

%\section{Time Series Recognition}

\begin{figure}[t]
\includegraphics[width=.85\columnwidth]{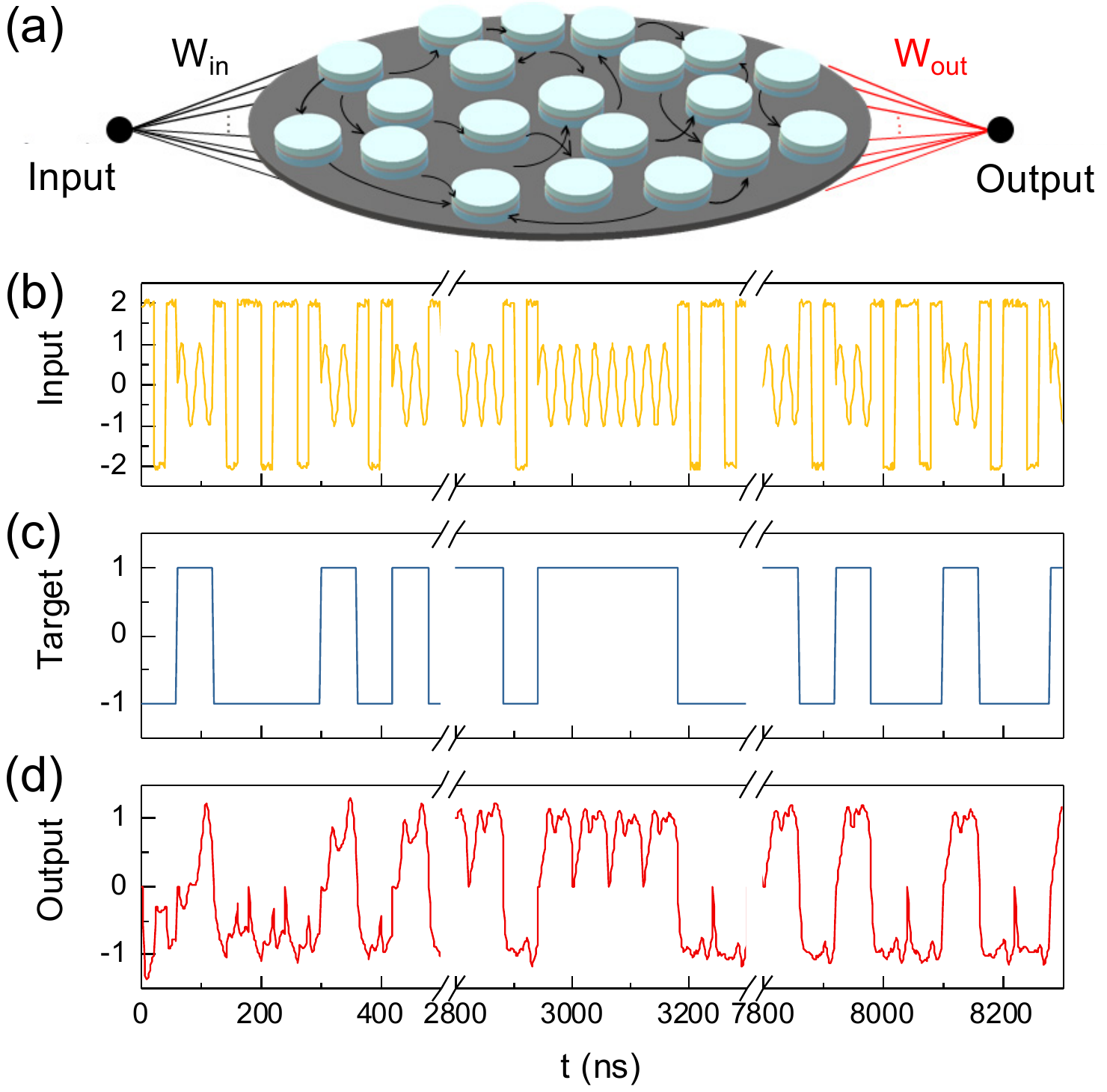}
\caption{(a) The structure of the RNN used for time series recognition. (b) The input data, (c) target function and (d) output of the RNN as a function of time. Learning is performed for the first 3600~ns, and the recognition is carried out in the next 6000~ns. All input functions have been successfully recognized.}
\label{fig:4}
\end{figure}
In addition to the generation of periodic functions, an RNN made of MTJs can also be applied to the recognition of time series. The structure of the network is shown in Fig.~\ref{fig:4}(a), where a time-dependent function is imported to the RNN from the input node. After adjusting the output weights $w_{\rm out}$ of the network, the output can have a different response to the corresponding input functions. To demonstrate recognition by the RNN, we input two simple functions into the RNN: a square wave $2\mathrm{sgn}[\sin(\omega_1 t)]$ and a sinusoidal function $\sin(\omega_2 t)$ with $\omega_1=0.16$~GHz and $\omega_2=0.21$~GHz. The designed target functions for the sinusoid and the square wave are $+1$ and $-1$, respectively.

For every 60~ns, we input one type of function, either the sinusoid or the square wave, as plotted in Fig.~\ref{fig:4}(b). Here, random noise is superposed on the function, which is approximately 5\% of the magnitude of the function. Then, we adjust the output weights $w_{\rm out}$ to let the output of the RNN match the required target function [Fig.~\ref{fig:4}(c)]. Such learning is performed 60 times in the first 3600~ns, and then the weights are fixed in later recognition. To avoid the influence of the previous recognition, we deliberately reset all the MTJs at the beginning of every recognition process, i.e., the initial output is always 0. The real RNN output is plotted in Fig.~\ref{fig:4}(d) as a function of time. Depending on the averaged output value, the RNN can recognize all the input waves 100 times (from 3600~ns to 9600~ns) successfully.

%\section{Discussions}

The numerical simulation we have done so far is a proof of concept and hence MTJs with fast precessions are chosen in simulation to reduce computational cost. In experiment lower-frequency precessions may be preferred, which can be done by using the vortex magnetization in the free layer \cite{Torrejon17}. Moreover, the high-frequency resistance is technically difficult to measure, so the measurable voltage of the MTJs can be used as the neuron output, which is just another nonlinear function of the input current. One also needs to consider two possible difficulties in experiment, i.e. the non-identical MTJs and phase noise, while the details are provided in Supplementary Material. We show that a RNN made of 40 MTJs with 25 different sizes works as well as the RNN with identical MTJs. Phase noise is one of the key issues limiting the functionality and performance of MTJ-based dynamical devices \cite{Chen16}. The RNN output is indeed affected by the phase noise, but can be systematically improved by increasing the number of MTJs in the RNN.

The synapses here are not realized using magnetic devices. Instead, we merely consider a hybrid system with the artificial neurons modeled by MTJs and an external storage for the synaptic weights. The implementation is analogous to the present neuromorphic chip, where static random access memory is employed to store the adjustable synaptic weights \cite{Merolla14,Akopyan15,Davies18,Pei19}. There are several proposals of trainable artificial synapses made of magnetic and resistive materials in literature \cite{Grollier16,Kurenkov19}, such as the Hall bars consisting of perpendicular magnetic multilayers \cite{Borders17}, the spintronics memristors based on domain walls \cite{Wang09,Sharad12,Lequeux16,Zhang19}, and the MTJ-based devices with multiple electrical resistances \cite{Krzysteczko12,Vincent15,Srinivasan16}. The technical challenge for applying these proposals is the effective and precise adjustment of the synaptic weights in the training process. The imperfection in synapses are explicitly examined in Supplementary Material including the fluctuation of the synaptic weights, signal delay in the RNN, and failure to update part of the output synapses during the training process. Nevertheless, the RNN can still learn to generate the target function indicating a high tolerance of the RNN for imperfect synapses.

%\section{Conclusions}

We have demonstrated that the generation and recognition of time series can be achieved by a MTJ-based recurrent neural network. Using micromagnetics to simulate the magnetization dynamics of the MTJs, we have shown that the RNN can learn to generate an arbitrary periodic function. With enough MTJs, an RNN can even be trained to simulate a handwritten character of the Chinese writing system. The recognition of different time-dependent functions has also been successfully performed using such a network. Moreover, this MTJ-based RNN is found to have a high tolerance to size dispersion of the MTJs. In time series recognition, such an RNN is resistant to the noise of the input signals.

The demonstration of this spintronic implementation of neuromorphic computing suggests that MTJs are very promising candidates for artificial neurons. Owing to the low energy cost and small geometric size, magnetic devices are expected to significantly improve the energy efficiency and integration density of neuromorphic devices. MTJs have ultrafast dynamics in the nanosecond regime and high endurance of more than $10^{15}$ cycles because magnetization dynamics does not involve any atomic motion as in the diffusive memristors. In addition, MTJs can be naturally integrated with the artificial synapses made of non-volatile magnetic memories such that all magnetic/spintronic neuromorphic chips can eventually be achieved. The proposed magnetic synapses also attract great attention in research \cite{Srinivasan16,Grollier16,Borders17,Kurenkov19,Wang09,Sharad12,Lequeux16,Zhang19,Krzysteczko12,Vincent15}, where the synaptic weight needs to be precisely adjusted during learning.

%%%%%%%%10%%%%%%%%20%%%%%%%%30%%%%%%%%40%%%%%%%%50%%%%%%%%60%%%%%%%%70%%%%%%%%80
\section*{Supplementary Material}
See supplementary material for the theoretical methods and numerical details, the effects of non-identical MTJs, phase noise of MTJs and the imperfect synapses.

\begin{acknowledgements}
This work was financially supported by the National Key Research and Development Program of China (Grant No. 2017YFA0303300), the National Natural Science Foundation of China (Grants No. 11734004, No. 61774018, No. 61604013, No. 61774017, and No. 31771146), the Recruitment Program of Global Youth Experts, and the Fundamental Research Funds for the Central Universities (Grants No. 2018EYT03 and No. 2018STUD03). Y.Y.M. acknowledges the financial support by Beijing Municipal Science and Technology Commission (Grant No. Z171100000117007) and Beijing Nova Program (Grant No. Z181100006218118).
\end{acknowledgements}

\end{document}